# Semi-Supervised Multi-Sequence Glioblastoma MRI Radiogenomics for Prediction of IDH Mutation Status: Improved Robustness to Limited Labels and SHAP Interpretations

Amir Hossein Pouria[1], Shahram Taeb[2], Somayeh Sadat Mehrnia[3⸸], Sajad Jabarzadeh Ghandilu[4⸸], Mehrdad Oveisi[5,6], Arman Rahmim[7,8], Mohammad R. Salmanpour[6,7,8*]

[1]Department of Computer Engineering, Amirkabir University of Technology, Tehran, Iran
[2]Department of Radiology, School of Paramedical Sciences, Guilan University of Medical Sciences, Rasht, Iran
[3]Department of Integrative Oncology, Breast Cancer Research Center, Motamed Cancer Institute, ACECR, Tehran, Iran
[4]Department of Electrical Engineering, Sharif University of Technology, Tehran, Iran
[5]Department of Computer Science, University of British Columbia, Vancouver, BC, Canada
[6]Technological Virtual Collaboration Company (TECVICO CORP.), Vancouver, BC, Canada
[7]Department of Radiology, The University of British Columbia, Vancouver, BC, Canada
[8]Department of Basic and Translational Research, BC Cancer Research Institute, Vancouver, BC, Canada

**(*) Corresponding Author:** Mohammad R. Salmanpour, msalman@bccrc.ca

**(⸸) Equal contribution as third co-author.**

## Abstract

**Background:** Glioblastoma (GBM) is an aggressive brain tumor, with IDH mutation status as a key prognostic biomarker. Traditional IDH testing requires invasive biopsies, highlighting the need for non-invasive alternatives. MRI-based radiogenomics features coupled with machine learning show promise, but past studies were mostly single-center-based and rarely used semi-supervised learning (SSL) to exploit unlabeled data.
**Methods:** We analyzed MRI sequences T1, T2-weighted, contrast-enhanced T1 (T1CE), and FLAIR from 1,329 patients across eight centers, with IDH labels available for 1,061 cases. Radiomics features (n=1,223 per case) were extracted using PyRadiomics with Laplacian of Gaussian and wavelet filters. Both supervised learning (SL) and SSL (via pseudo-labeling) were implemented, incorporating 38 feature selection/attribute extraction and 24 classifiers. Five-fold cross-validation was performed on UCSF-PDGM and UPENN datasets, with external validation on IvyGAP, TCGA-LGG, and TCGA-GBM. SHAP analysis quantified feature importance.
**Results:** Multimodal MRI (T1+T2+T1CE+FLAIR) provided the strongest performance, outperforming single-sequence models. The best SSL model (involving Recursive Feature Elimination (RFE) + SVM) achieved 0.93±0.01 cross-validation and 0.75±0.02 external accuracy, while the best SL model (RFE+Complement Naive Bayes (CNB)) reached 0.90±0.02 and 0.80±0.006, respectively. SSL also demonstrated greater stability with lower sensitivity to dataset size compared to SL, maintaining robust performance in data-limited conditions. SHAP analysis showed SSL amplified the discriminative value of first-order statistics of Root Mean Square (FO_RMS) (T1CE) and wavelet-based metrics, strengthening biomarker interpretability.
**Conclusion:** SSL improves accuracy, efficiency, and interpretability in MRI-based IDH prediction, remaining less sensitive to data size while reinforcing multimodal fusion as the most reliable, scalable strategy.

## 1. Introduction

Glioblastoma (GBM) is the most common and aggressive malignant brain tumor, representing the highest-grade end of a heterogeneous glioma spectrum [1]. Accurate classification is central to clinical decision-making because it informs prognosis and therapy [2]. Among molecular biomarkers, isocitrate dehydrogenase (IDH) mutation status is a critical determinant of outcome: IDH-mutant gliomas generally exhibit longer survival, whereas IDH wild-type tumors—often GBM—follow a more aggressive course [3]. Crucially, early and reliable prediction of IDH mutation status affects diagnosis, treatment selection, and patient counseling, enabling personalized care before histopathology is available [4].

In current practice, IDH mutation status is specified by histopathology and genomic testing under the World Health Organization (WHO) classification, but these require invasive biopsy and timely access to molecular assays—constraints that are not universally met [5]. A non-invasive imaging-based alternative can accelerate risk stratification and treatment planning, especially when surgery is contraindicated or molecular testing is delayed or unavailable [6]. Magnetic resonance imaging (MRI) is a cornerstone of glioma evaluation, and radiogenomics analyses enable quantitative analysis of tumor characteristics—shape, intensity, texture, and heterogeneity—providing objective imaging biomarkers that can support or, in select contexts, substitute for invasive testing [7, 8].

Some studies [9, 10, 11] have explored MRI-based prediction of IDH mutation status using radiogenomics feature (RFs) and machine learning. Zhang et al. [12] combined deep learning signatures and conventional RFs across T1-weighted (T1), T2-weighted (T2), contrast-enhanced T1 (T1CE), and Fluid-Attenuated Inversion Recovery (FLAIR) sequences to achieve high accuracy in IDH genotyping, demonstrating the value of multimodal data fusion. Multiparametric radiomic models, incorporating T1, T2, T1CE, and FLAIR, have also shown excellent performance, supporting the power of feature integration across sequences [13]. Deep learning approaches further enhance accuracy: Yan et al. demonstrated that deep learning features derived from diffusion tensor imaging (DTI) improve glioma molecular stratification [14], while Pasquini et al. applied CNNs on multiparametric MRI in GBM and achieved high accuracy [15]. Although deep RFs often outperform handcrafted ones, they are less reproducible because they are data-driven, architecture-dependent, and sensitive to acquisition or preprocessing variations, which limits their stability across centers and hinders clinical translation [16, 17, 18]. Despite these advances, many prior studies were conducted in single-center cohorts with limited sample sizes, raising concerns about generalizability. Few systematically compared the relative contribution of individual MRI sequences versus their combinations, and almost none incorporated semi-supervised learning (SSL) to leverage unlabeled data. Moreover, feature importance analyses—to enhance interpretability and reproducibility—remain underutilized.

Despite substantial progress with machine learning (ML) and RFs, uncertainty persists regarding the most informative MRI sequence(s) for predicting IDH mutation status [8]. This inconsistency motivates multimodal strategies that integrate complementary information across sequences—vascular enhancement on T1CE, edema and tissue water on T2/FLAIR, and structural detail on T1—to capture a more complete picture of tumor biology [19]. In parallel, deep learning automates hierarchical feature discovery from images, while quantitative analysis standardizes heterogeneous MRI inputs into comparable representations, creating a robust and generalizable substrate for multimodal fusion and cross-site comparison [17].

For clinical translation, interpretability is essential. Assessing feature importance and contribution strengthens generalizability by emphasizing stable, biologically plausible predictors; improves reproducibility by focusing on features reliably extracted across scanners, protocols, and preprocessing; and enhances clinical trust by linking influential imaging patterns to known pathophysiology. Feature-driven model refinement also reduces dimensionality and mitigates overfitting, yielding leaner, more stable predictors [20, 21]

Two practical barriers limit deployment: data scarcity and multicenter variability. Labeled medical imaging data are costly to obtain, and models trained on single-center cohorts often degrade on external data due to scanner and protocol differences. SSL addresses both challenges by leveraging abundant unlabeled cases alongside limited labeled data, effectively expanding training size and improving robustness [22]. When combined with multimodal fusion and quantitative harmonization [23, 24], SSL can counteract site-specific biases and support clinically meaningful generalization [22].

This study is important because it targets these translational bottlenecks—reliance on single sequences, small single-center datasets, limited labels, and insufficient interpretability—within a unified framework. Our contributions are threefold: (1) a systematic, multicenter evaluation of individual and combined MRI sequences for prediction of IDH mutation status using radiogenomics analyses; (2) an integrated SSL and supervised learning (SL) pipeline that leverages unlabeled data to increase effective sample size; and (3) a feature-importance analysis to enhance interpretability, generalizability, and reproducibility across sites. Together, these elements advance a scalable, transparent approach to imaging-based molecular prediction in neuro-oncology.

## 2. Materials and Methods

### 2.1 Patient Data

We collected data from 1,329 GBM patients across 8 centers [ACRIN-FMISO-Brain (# of 4) [25], Brain-Tumor-Progression CPTAC-GBM (# of 33) [26], IvyGAP (# of 30) [27], REMBRANDT (# of 63) [28], TCGA-GBM (# of 167) [29], TCGA-LGG (# of 263) [30], UCSF-PDGM (# of 202) [31], UPENN-GBM (# of 567) [32] with clinical data, delineated masks, and various MRI sequences (T1, T2, T1CE, and FLAIR) from The Cancer Imaging Archive (TCIA). All MRI images were reviewed to ensure high-quality, artifact-free data. These images were then enhanced and normalized. Our study aimed to improve the prediction of IDH mutation status in GBM by developing two frameworks: SL and SSL strategies. A total of 1,223 RFs were extracted from the MRI sequences using Laplacian of Gaussian (LoG; $\sigma$ = 1.0, 2.0, 3.0, 4.0, and 5.0 mm) and wavelet (LLH, LHL, LHH, HLL, HLH, HHL, HHH, LLL) filters, each applied with varying parameter settings to capture a broad range of spatial and textural characteristics. The extracted features were normalized using min–max scaling. From these RFs, fifteen combined datasets were generated, each representing different combinations of MRI-derived features. Among the 1,329 patients included, IDH mutation status was available for 1,061 patients, while the remaining patients lacked outcome data.

Demographic and clinical characteristics varied across datasets. For instance, the ACRIN 6684 dataset includes 45 patients with newly diagnosed GBM multiforme who underwent baseline MRI, 18F-FMISO PET, and low-dose Computed Tomography (CT) imaging. The cohort had a mean age of 57.2 years (range 29–77), with 64% male and 36% female; most were White (91.1%), with smaller proportions of Black, Asian, and American Indian/Alaska Native patients. In addition to imaging, the dataset provides clinical, demographic, treatment, and biomarker data (e.g., MGMT, HIF1-α, GLUT1, CAIX), enabling integrative analyses of tumor hypoxia and therapeutic response.

Each of these landmark glioma datasets offers distinct imaging characteristics that enrich radiogenomic research: CPTAC-GBM integrates multimodal imaging (MRI, CT, histopathology) with high-resolution whole-slide data to align imaging features with proteogenomics; IvyGAP provides serial MRI (pre-, post-, and follow-up) with detailed histologic annotation, enabling spatially resolved correlations of contrast-enhanced tumor regions with gene expression; REMBRANDT contributes multi-sequence MRI of gliomas with variable resolution, complemented by molecular and clinical data for prognostic modeling; TCGA-GBM and TCGA-LGG aggregate MRI and CT acquired across diverse scanners and institutions, offering heterogeneous but highly representative cohorts for studying resolution and contrast variability in relation to genomic drivers; UCSF-PDGM stands out with standardized 3T MRI protocols including advanced diffusion (HARDI) and perfusion (ASL) imaging, providing higher spatial and functional resolution alongside expert tumor segmentations; and UPENN-GBM delivers the largest mpMRI collection with co-registered, segmented volumes and curated RFs, optimized for reproducible AI/ML applications. Together, these datasets capture a spectrum of imaging resolutions, contrasts, scanner heterogeneity, and advanced protocols, providing complementary strengths for radiogenomic discovery and precision neuro-oncology. Only MRI sequences from each dataset were used in this study.

### 2.1. Classification Analysis:

As depicted in Figure 1, the proposed pipeline offers a comprehensive framework for constructing robust ML models using RFs derived from the brain MRI of IDH patients. The pipeline includes image preprocessing, dimension reduction via feature selection algorithms (FSAs) and attribute extraction algorithms (AEAs), classifier benchmarking, and thorough validation under both SL and SSL frameworks.

**i) Mask Validation and ii) Expert Verification.** Brain MRI examinations, encompassing T1, T2, T1CE, and FLAIR sequences, were thoroughly evaluated to detect glioma-related abnormalities. A dual review by two board-certified neuroradiologists standardized tumor annotations, improved inter-observer reliability, and ensured accurate localization. Cases with unclear tumor boundaries—due to motion artifacts, hemorrhage, or significant postoperative changes—were excluded from the analysis.

**iii) MRI Intensity Normalization.** To address variations in MRI acquisition parameters and patient anatomy, each MRI sequence was subjected to intensity normalization using the min–max method. This process rescales voxel intensity values to a standardized range, typically (0, 1), ensuring stable input for RFs extraction. Normalization also facilitates the comparability of intensity-dependent features across patients, sequences, and imaging centers.

**iv) RF Extraction.** After normalization, brain MRI sequences were processed using the open-source PyRadiomics package, compliant with the Image Biomarker Standardization Initiative (IBSI) [33] guidelines to ensure reproducibility and consistency. PyRadiomics facilitated the extraction of a comprehensive set of 107 RFs from each sequence, capturing diverse properties such as morphology and microstructural organization, which are essential for glioma subtyping and IDH mutation prediction. The extracted features comprised 19 first-order statistics (FO), 16 three-dimensional shape features, 10 two-dimensional shape features, 23 gray level co-occurrence matrix (GLCM) features, 16 gray level size zone matrix (GLSZM) features, 16 gray level run length matrix (GLRLM) features, 5 neighboring gray tone difference matrix (NGTDM) features, and 14 gray level dependence matrix (GLDM) features. These standardized features, derived from normalized MRI volumes, are designed to capture both global and local imaging patterns critical for IDH prediction in GBM characterization.

**v) Data Splitting and vi) Normalization Strategy.** After feature extraction, the datasets were split into a five-fold cross-validation set and distinct external testing sets. The UCSF-PDGM and UPENN-GBM datasets, which include IDH mutation data, were used as the primary training cohort and subjected to five-fold cross-validation. To assess model generalizability, three independent datasets with IDH mutation—IvyGAP, TCGA-LGG, and TCGA-GBM—each from different clinical centers, were designated for external testing. Datasets without outcome information were excluded from supervised training and instead utilized in the SSL process. To maintain methodological rigor and prevent data leakage, normalization parameters (e.g., min and max) were calculated solely from the training folds (four divisions) and then applied to the validation fold, unlabeled datasets, and external test sets during evaluation.

**vii) SL approaches:** In the SL framework, the labeled UCSF-PDGM and UPENN-GBM datasets were divided into five folds. For each iteration, four folds were utilized for training, with the remaining fold reserved for validation, ensuring each fold served as the validation set once across five iterations to complete the cross-validation cycle. Additionally, the model trained on each fold was evaluated on three independent labeled datasets—IvyGAP, TCGA-LGG, and TCGA-GBM—to assess its generalizability across diverse centers and patient cohorts. Performance metrics, including Accuracy, Precision, Recall, F1-score, Receiver Operating Characteristic – Area Under the Curve (ROC-AUC), and Specificity [34]— were reported as average values with standard deviations across the five cross-validation folds and external test evaluations. Model selection was determined by the highest performance across all metrics during five-fold cross-validation, with external validation conducted using independent test sets.

**viii) SSL approaches:** Within the SSL framework, the labeled UCSF-PDGM and UPENN-GBM datasets were divided into five folds. In each iteration, a logistic regression (LR) model was trained on four labeled folds and subsequently used to assign pseudo-labels to the unlabeled datasets (e.g., ACRIN-FMISO-Brain, CPTAC-GBM, and REMBRANDT). To avoid bias and data leakage, the remaining labeled fold was excluded from the pseudo-labeling process. Following pseudo-labeling, the model was retrained using the combined labeled and pseudo-labeled data from the four folds and evaluated on the held-out validation fold and three external test sets—IvyGAP, TCGA-LGG, and TCGA-GBM—to determine the contribution of unlabeled data to enhancing model generalization.

**ix) Dimensionality Reduction via FSAs and AEAs.** To mitigate the high dimension of RFs and minimize overfitting risks, our pipeline employs two parallel approaches: FSAs and AEAs [35]. We evaluated 38 dimensionality reduction techniques (19 FSAs and 19 AEAs) for their effectiveness in identifying the most informative and non-redundant features. The 19 FSAs are categorized into three main groups. Filter-based methods, including Chi-Square Test (CST), Correlation Coefficient (CC), Mutual Information (MI), and Information Gain Ratio (IG), score features independently of classifiers. Statistical tests such as ANOVA F-Test (AnovaFT), ANOVA P-value selection, Chi2 P-value, and Variance Thresholding (VT) evaluate feature discriminativeness. Wrapper-based methods, such as Recursive Feature Elimination (RFE), Univariate Feature Selection (UFS), Sequential Forward Selection (SFS), and Sequential Backward Selection (SBS), iteratively assess feature subsets based on model performance. Embedded methods, including Lasso, Elastic Net (ENet), Embedded Elastic Net (EmbENet), and Stability Selection, integrate FSAs into the training process. Ensemble-based methods like Feature Importance by RandF (FIRF), Extra Trees (ETI), and Permutation Importance (Perm-Imp) capture complex nonlinear relationships. Additional statistical controls, such as False Discovery Rate (FDR), Family-Wise Error (FWE), and multicollinearity handling via Variance Inflation Factor (VIF), are also applied. Dictionary-based strategies leverage Principal Component Analysis (PCA) or sparse loadings for enhanced stability and interpretability. Features selected by FSAs are detailed in the Supplementary Files 1-10 (ten files for each sequence and the multiparametric set) for both SL and SSL frameworks.

AEAs provide a complementary approach by projecting the feature space into lower-dimensional subspaces. The 19 AEAs include linear methods like PCA, Truncated PCA, Sparse PCA (SPCA), and Kernel PCA (with RBF and polynomial kernels), which identify uncorrelated projections with maximum variance. Independent Component Analysis (ICA) isolates statistically independent latent variables, while Factor Analysis reveals underlying structures in observed features. Non-negative Matrix Factorization (NMF) produces interpretable parts-based decompositions. SL methods like Linear Discriminant Analysis (LDA) optimize class separation in the transformed space. Advanced manifold learning techniques, including t-SNE, Uniform Manifold Approximation and Projection (UMAP), Isomap, Locally Linear Embedding (LLE), Spectral Embedding, Multidimensional Scaling (MDS), and Diffusion Maps, capture non-linear structures, aiding visualization of complex relationships in high-dimensional radiogenomics data. Deep learning approaches, such as shallow and deep autoencoders, facilitate data-driven feature compression via reconstruction optimization. Additional methods include Feature Agglomeration for hierarchical clustering, Truncated SVD for matrix decomposition, and projection-based techniques like Gaussian Random Projection, Sparse Random Projection, and Feature Hashing, which offer scalable compression solutions.

**x) Classification Algorithms.** Each reduced feature set, whether obtained through FSAs or AEAs, was assessed using a comprehensive set of 24 classifiers (CAs). These encompassed linear models, tree-based classifiers such as Decision Trees (DT), Random Forest (RandF), Extra Trees (ET), Gradient Boosting (GB), AdaBoost (AB), and HistGradient Boosting (HGB), which leverage ensemble learning to minimize variance and enhance generalization. Meta-ensemble approaches, including Stacking, Voting Classifiers (VC), and Bagging, further improved predictive robustness by combining the strengths of multiple base learners. Support Vector Machines (SVM) were applied with various kernels to address both linear and non-linear classification tasks, while k-Nearest Neighbors (KNN) offered a distance-based, instance-level method. Several Naive Bayes variants, including Gaussian (GNB), Bernoulli (BNB), and Complement Naive Bayes (CNB), were evaluated for their probabilistic simplicity and computational efficiency. The Multi-Layer Perceptron (MLP), a neural network-based approach, enabled modeling of complex, non-linear patterns, while gradient-boosted frameworks like Light GBM (LGBM) and XGBoost (XGB) delivered high-performance learning through gradient optimization and feature importance analysis. Additional classifiers included Linear Discriminant Analysis (LDA), Nearest Centroid (NC), Decision Stump, Dummy Classifier (DC), Gaussian Process Classifier (GP), and Stochastic Gradient Descent Classifier (SGDC), providing a range of modeling strategies. All classification algorithms were optimized using five-fold cross-validation and grid search. The optimal hyperparameters for each model in both SL and SSL frameworks are documented in Supplementary Files 1 and 10, Sheet 4.

**xi) Assess Sensitivity of Top-Performing Models to Data Size in SL and SSL.** To examine how data volume influences model performance under SL and SSL, we designed three experiments, each repeated with 100 randomized arrangements (bootstraps). In Scenario 1 (SL), training began with 10% of the labeled UCSF-PDGM and UPENN-GBM datasets, then increased in 10% increments until 100% was used, enabling assessment of performance gains with more labeled data. In Scenario 2 (SSL), both labeled and unlabeled data were expanded together from 10% to 100% in 10% steps to test the effect of simultaneous growth. In Scenario 3 (SSL), the full labeled dataset was fixed while the unlabeled pool was gradually added from 10% to 100%, isolating the contribution of unlabeled data. This setup investigated the contribution of additional unlabeled data to model performance with a fixed labeled dataset. Collectively, these experiments offered a thorough understanding of model robustness and adaptability to varying data volumes in SL and SSL contexts. Across all scenarios, only external testing metrics were reported, as internal training and validation splits varied dynamically with data volume changes. Using fixed external test sets ensured consistent and equitable model comparisons across different data conditions.

**xii) Feature Importance Investigation by SHAP.** To examine and elucidate the role of individual RFs in classification results, we utilized SHapley Additive exPlanations (SHAP) on 25 high-performing combinations of ML models with FSAs or AEAs, selected for their outstanding accuracy in predictive performance. For each combination, we calculated SHAP values to assess the marginal impact of each feature on the model's predictions, distinguishing between class 0 (wild-type IDH) and class 1 (mutant IDH) cases. These SHAP values were then averaged across all combinations within each class to provide a more robust and comprehensive view of feature importance trends. The averaged SHAP values were visualized using heatmaps, facilitating a comparative analysis of feature contributions across the two classes. This methodology improves model transparency and interpretability while strengthening the biological plausibility and diagnostic relevance of the selected features for IDH prediction in GBM characterization.

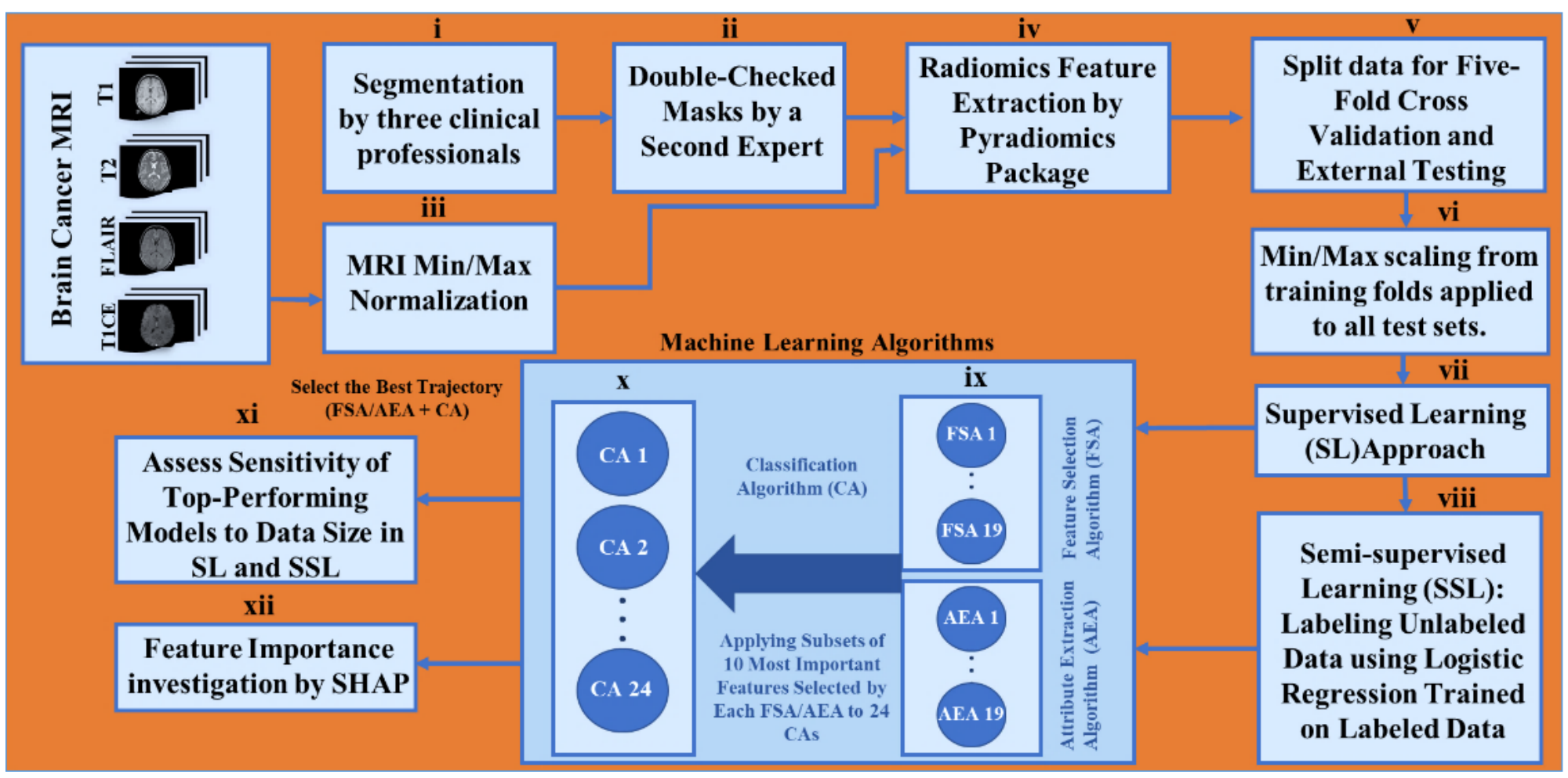

**Fig 1.** Integrated Radiomics-Based ML Pipeline for Brain Cancer Classification and Prognosis. The pipeline integrates expert-guided tumor segmentation, min-max MRI normalization, IBSI-compliant RFs extraction, dimensionality reduction, and classification algorithms. SL and SSL strategies are implemented with five-fold cross-validation and external validation. Model robustness is evaluated through sensitivity analysis to data size, and feature importance is interpreted using SHAP. Abbreviations: ML: Machine Learning, SL: Supervised Learning, SSL: Semi-supervised learning, MRI: Magnetic Resonance Imaging, IBSI: Image Biomarker Standardization Initiative, RFs: Radiogenomics Features, SHAP: SHapley Additive exPlanations.

## 3. Results

### 3.1. Classification Analysis Result

Figure 2 illustrates the comparative performance of different dimensionality reduction–classifier combinations across individual MRI sequences (T1, T2, T1CE, and FLAIR) as well as their combined use. While accuracy and its standard errors remain central, additional evaluation metrics such as F1 score, AUC, precision, recall, and specificity provide a more nuanced picture of each model's behavior.

Within the SSL framework, the combination of RFE with SVM on all four MRI sequences (T1, T2, T1CE, FLAIR) consistently outperformed other pipelines. Beyond its strong cross-validation and test accuracies ($0.93 \pm 0.01$ and 0.75 ± 0.02, respectively), this model achieved a high F1 score (0.94 validation, 0.74 test) and excellent discriminative ability as indicated by its AUC (0.97 validation, 0.84 test). Specificity also remained robust (0.93 validation, 0.92 test), underscoring its balance between sensitivity and reliability in excluding false positives. A close competitor was RFE with CNB on the same combined dataset, which, although slightly lower in overall accuracy (0.87 ± 0.02 CV, 0.72 ± 0.03 test), maintained competitive F1 performance (0.89 validation, 0.70 test) and respectable AUC values (0.93 validation, 0.78 test). By contrast, alternative SSL pipelines such as CC with VC or t-SNE with GNB displayed lower stability, reflected in F1 scores near 0.58–0.73 and notably reduced AUC values (as low as 0.59 in validation and 0.58 in testing), confirming the superiority of RFE-based methods in capturing informative features. Under the SL framework, performance trends were largely consistent. RFE with CNB on the multi-sequence mixture not only delivered the highest test accuracy (0.80 ± 0.006) but also maintained excellent precision, recall, and F1 balance (all at 0.80). Its AUC of 0.96 (validation) and 0.86 (test) confirmed strong discriminatory power. Similarly, RFE with SVM achieved a validation F1 score of 0.92 and a test F1 of 0.78, with AUC values of 0.96 and 0.85, respectively. Importantly, the exceptionally small test error for RFE–CNB demonstrates superior robustness to unseen cases. Other SL methods, such as CC–VC and t-SNE–GNB, again showed substantially weaker performance, with test F1 scores around 0.51–0.62 and lower AUC values (0.62–0.69).

Looking at individual modalities, T1CE emerged as the strongest single sequence. Under SSL with AE–LR, it achieved validation F1 and AUC values of 0.93 and 0.94, respectively, with external test F1 and AUC still competitive at 0.74 and 0.82. Other T1CE-based pipelines, including ETI–SVM and MI–GNB, further demonstrated reliable generalization, with test F1 values in the 0.77–0.81 range and AUC spanning 0.70–0.85. In contrast, T1 and T2 modalities exhibited moderate results. For instance, T1 with SSL–AnovaFT–GNB achieved validation F1 of 0.87 but dropped to 0.69 on the test set; T2 pipelines generally peaked at validation F1 ~0.85 but consistently dropped below 0.65 on external evaluation, reflecting limited robustness when used in isolation. FLAIR alone proved the weakest, with best-performing SSL–EmbENet–LR reaching validation F1 of 0.83 but only 0.59 in test F1, accompanied by reduced AUC values (0.88 validation, 0.65 test). This highlights the instability of FLAIR-derived models relative to multi-sequence integration.

Taken together, the results show that incorporating multi-sequence input with RFE-based FSAs and robust classifiers (SVM, CNB) yield the most balanced trade-offs across accuracy, F1, and AUC, outperforming individual modalities and alternative selectors such as LASSO, UMAP, or t-SNE. Notably, while SSL models provided slight gains in validation accuracy and F1, SL models matched or exceeded their performance in test generalization, particularly when stability and specificity were considered. Statistical analysis confirmed significant differences ($p < 0.05$, Benjamini–Hochberg corrected) between top-performing pipelines on T1CE and combined sequences compared to weaker single modalities such as FLAIR or T2. A comprehensive listing of all results, including feature sets and hyperparameters, is available in Supplemental Files 1–10 for SL and SSL analyses.

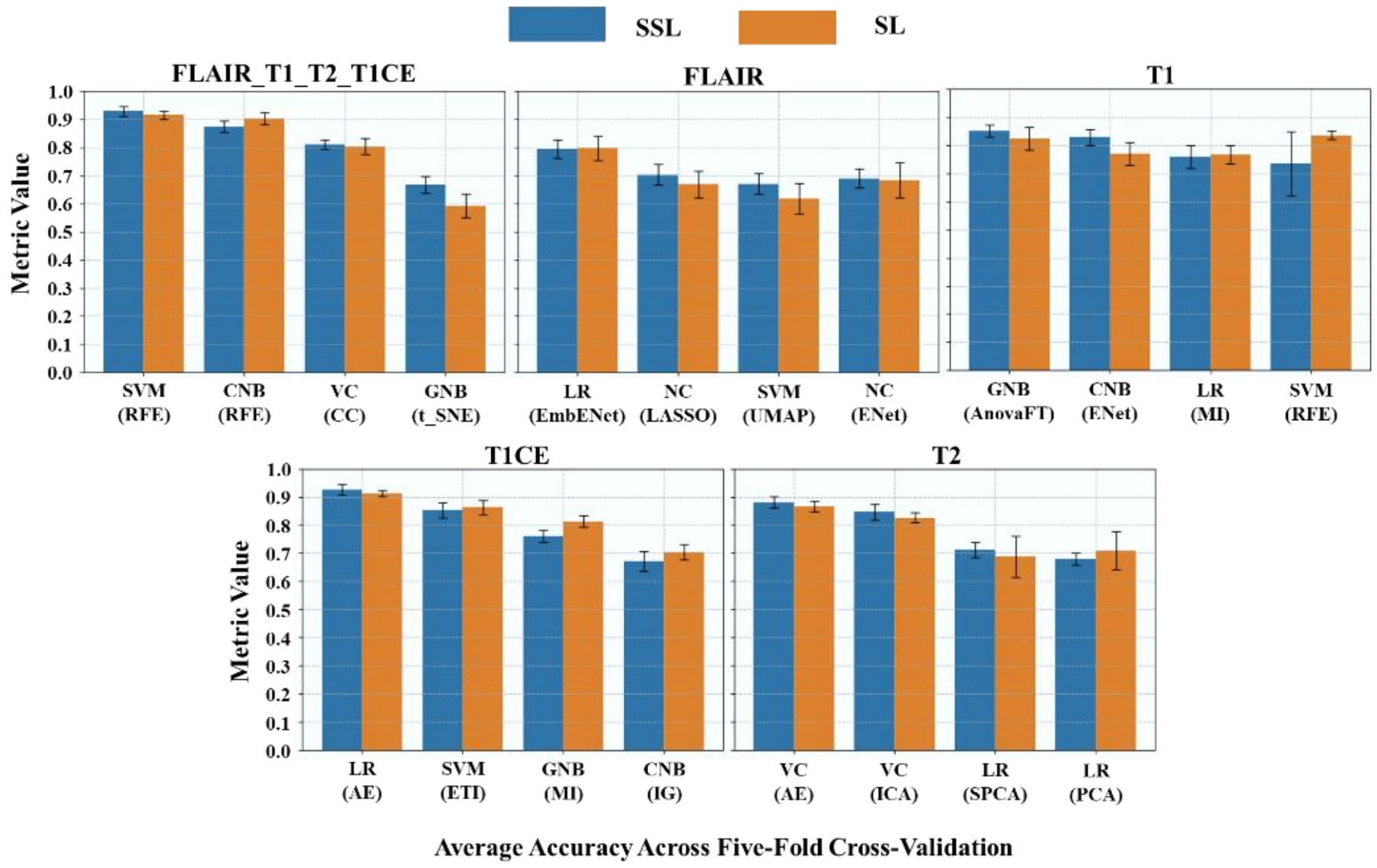

**Fig 2**. Bar plots showing the average accuracy of different ML techniques applied to various combinations of RFs extracted from different MRI sequences. The X axis indicates the best ML techniques: Classifiers + Dimension Reduction algorithms, and the Y axis shows the metric values of average accuracy across five-fold cross-validation.

### 3.2. Impact of Data Size on the Sensitivity of SL and SSL Models

In the first scenario, the optimal SL configuration (Complement Naïve Bayes + RFE) was trained on progressively larger fractions of labeled data (10%–100%), while the external test sets (IvyGAP, TCGA-LGG, and TCGA-GBM) remained fixed. Averaged over 100 random splits of the training data, where models were retrained and evaluated on the same fixed external test sets, performance decreased slightly from 0.92 with 10% labeled data to 0.90 with the full dataset, indicating diminishing returns from additional labeled samples. Interestingly, this decline was not uniform across all evaluation criteria. Precision remained consistently high (~0.93–0.95), showing the model's robustness in minimizing false positives. Recall and F1-score, however, showed modest fluctuations (~0.89–0.92 and ~0.90–0.91, respectively), reflecting slight trade-offs between sensitivity and overall balance of predictions. AUC values remained strong (~0.94–0.96), demonstrating stable discriminative ability, while specificity varied more widely (~0.76–0.82 on test sets), suggesting that SL models were more sensitive to negative class misclassification as dataset size increased. In the second scenario, when both labeled and unlabeled samples were increased simultaneously in 10% increments, the SSL model (SVM + RFE) improved steadily, rising from 0.87 with 10% data to 0.93 with the full dataset. Beyond 60%, accuracy gains tapered, but additional metrics highlighted SSL's advantage. F1-scores increased in parallel with accuracy (from ~0.87 to ~0.94), indicating that improvements were not biased toward precision or recall alone but rather enhanced the balance between them. AUC values remained consistently higher than SL (~0.95–0.97), confirming that SSL models better captured class separability. Moreover, specificity reached up to 0.92 on external test sets, markedly outperforming SL, suggesting that SSL approaches were more reliable at correctly identifying negative cases.

In the third scenario, the same SSL configuration (SVM + RFE) was evaluated by fixing the labeled set and incrementally adding unlabeled samples. Accuracy rose quickly to 0.91 with only 10% unlabeled data and stabilized, peaking at 0.93 with the full pool. Importantly, precision and recall converged to ~0.93–0.95, resulting in F1-scores exceeding 0.92 across most runs, indicating a consistent balance. AUC remained high (~0.95–0.97), reinforcing SSL's robustness. Specificity trends highlighted one of the most notable advantages of SSL: with even small additions of unlabeled data, specificity quickly increased above 0.90 and remained stable, in contrast to SL, where variability persisted.

Taken together, these findings indicate that while SL models achieved strong precision, their recall, F1, and specificity were more sensitive to dataset size, leading to variability in generalization. By contrast, SSL models consistently maintained high and stable F1 and AUC values alongside improved specificity, demonstrating their ability to balance positive and negative case detection while reducing overfitting to limited labeled data. Across all scenarios, SSL approaches (particularly Scenario 3) consistently outperformed SL in data-limited contexts, underscoring the value of leveraging unlabeled data to improve generalization and reduce sensitivity to dataset size (Figure 3).

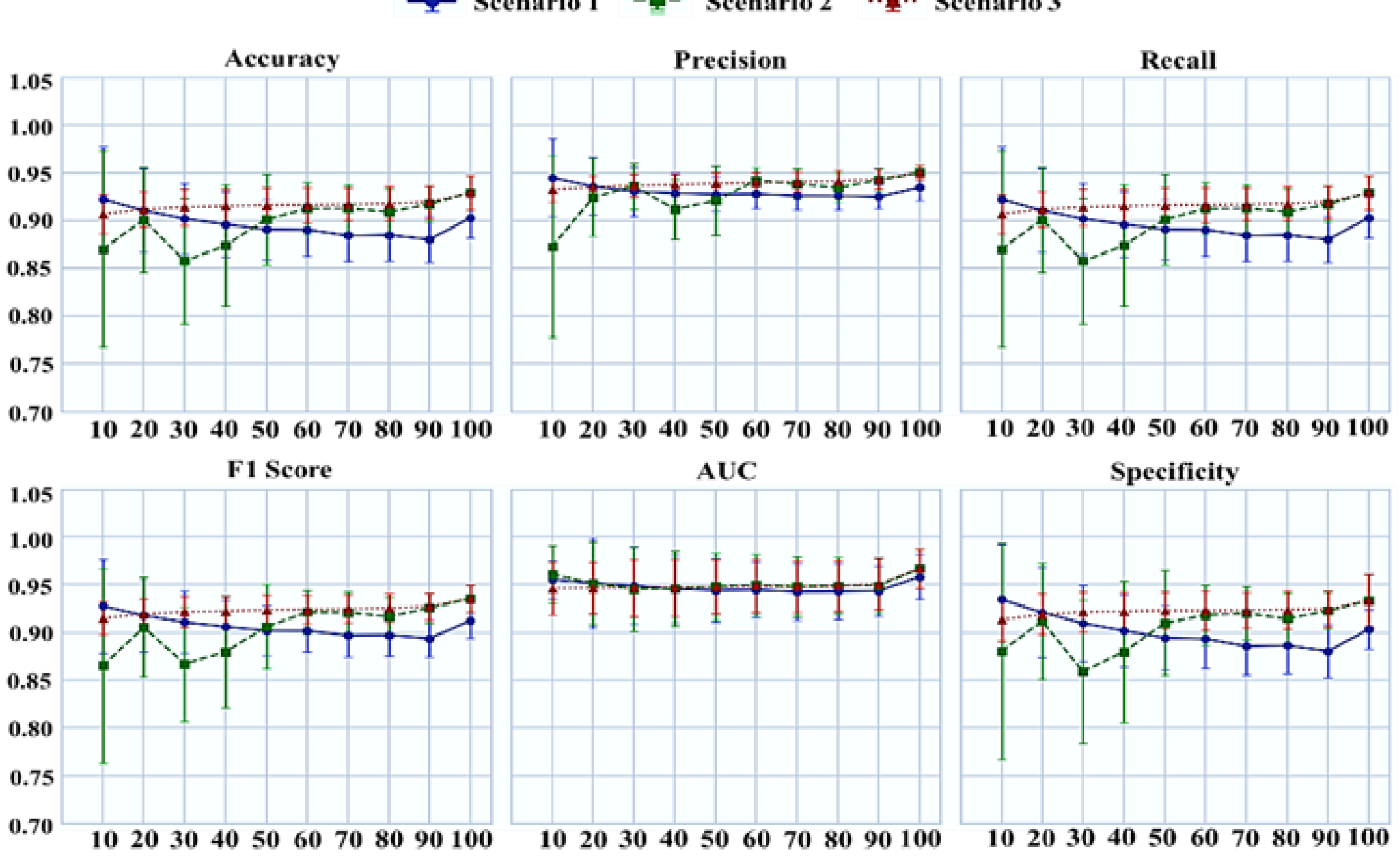

**Fig. 3.** Curve plots illustrate the sensitivity of top-performing SL and SSL models for IDH prediction with respect to data size. Average 5-fold cross-validation metrics (F1-score, AUC, specificity, accuracy, precision, and recall) are reported for three scenarios: Scenario 1 (SL: RFE_CNB), Scenario 2 (SSL + SL: RFE_SVM), and Scenario 3 (SSL: RFE_SVM). Performance is assessed across varying proportions of training data (10%–100%). Abbreviations: SL, supervised learning; SSL, semi-supervised learning; IDH: Isocitrate Dehydrogenase, AUC: Area Under the Curve, RFE: Recursive Feature Elimination, CNB: Complement Naive Bayes, SVM: Support Vector Machine,

### 3.3. SHAP-Based Feature Importance in SL and SSL for GBM IDH Mutation Prediction

In this section, model interpretability was investigated using SHAP to evaluate feature importance in SL and SSL frameworks. The analysis focused on the top 25 high-performing combinations of ML classifiers and FSAs or AEAs, identified through five-fold cross-validation accuracy. These combinations consisted of five distinct classifiers—Random Forest, Support Vector Machine, Gradient Boosting, LR, and Multi-Layer Perceptron—paired with five FSAs or AEAs, applied to RFs extracted from multi-center MRI sequences.

In the SL setting, Fig. 4(i) presents a heatmap visualizing the average SHAP values for class 0 (wild-type IDH) and class 1 (mutant IDH) across features selected by the top 25 model-FSA/AEAs combinations in a binary classification task. Each row corresponds to a specific RF, such as textural (e.g., Gray Level Co-occurrence Matrix, GLCM), morphological, or wavelet-transformed features, while the two columns represent the average absolute SHAP values for class 0 and class 1, respectively. The heatmap, Fig. 4, employs a color gradient to highlight relative feature importance: deep red indicates features with stronger contributions to class 0 predictions, while deep blue signifies greater influence on class 1 predictions. Class 1 (Mutant IDH): Features such as Sphericity (SF_Sp_3D, original, FLAIR), Difference Entropy (GLCM_DiEn, LoG sigma:2.0, T1CE), Informational Measure of Correlation (GLCM_IMC2, wavelet LHL, T1CE) exhibited strong positive SHAP contributions, indicating their critical role in predicting IDH mutation status. These features are associated with higher intensity blue regions in the heatmap, reflecting their robust influence. In addition, Class 0 (Wild-type IDH): Features like Contrast (NGTDM_con, wavelet LHL, T2), Root Mean Square (FO_RMS, original, T1CE), Inverse Difference Normalized (GLCM_IDN, wavelet LLH, T1CE) showed moderate positive SHAP contributions, with less intense red regions compared to class 1 features. This suggests a relatively weaker but still notable influence on wild-type IDH predictions. In conclusion, Class 1 features generally demonstrated stronger individual contributions, as evidenced by the more pronounced blue regions in the heatmap. This facilitates both model interpretability and the identification of potential biomarkers for IDH mutation prediction in GBM.

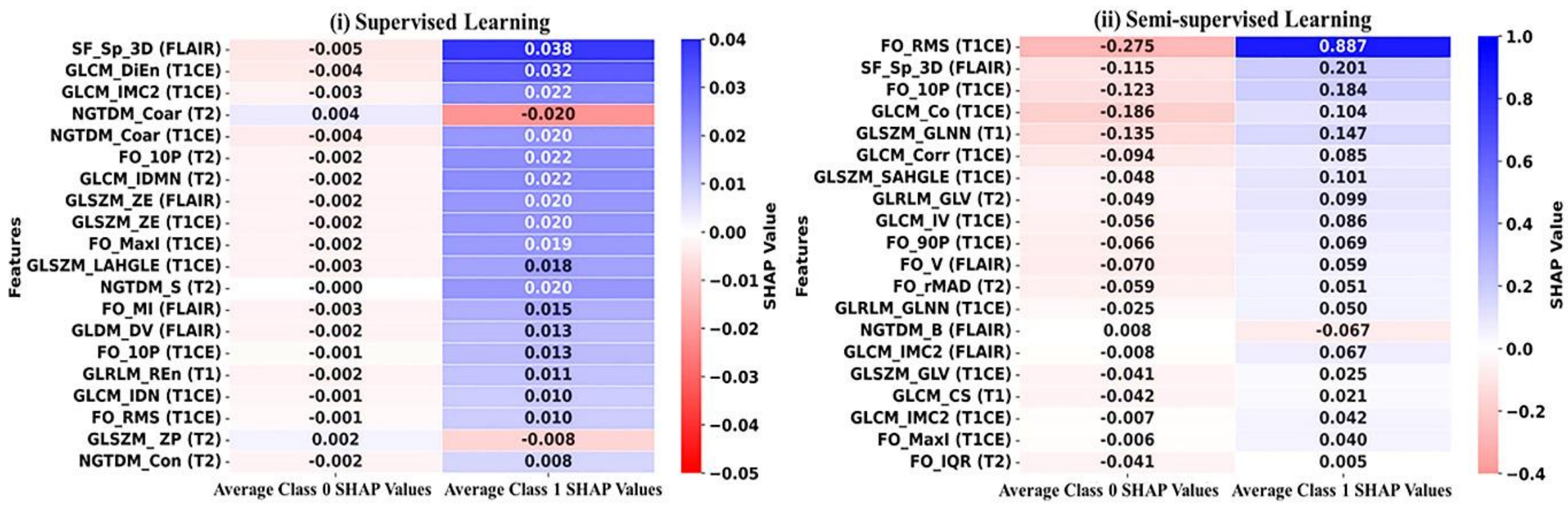

**Fig. 4.** Heatmaps illustrate the average SHAP values for feature importance in (i) SL and (ii) SSL settings for predicting IDH mutation status in GBM. Red tones indicate stronger contributions to class 0 (wild-type IDH), while blue tones highlight features with greater influence on class 1 (mutant IDH). The following RFs are included: SF_Sp_3D: Sphericity (Sp), GLCM_DiEn: Difference Entropy (DiEn), GLCM_IMC2: Informational Measure of Correlation (IMC2), NGTDM_Coar: Coarseness (Coar), FO_10P: The 10th Percentile (10P), GLCM_IDMN: Inverse Difference Moment Normalized (IDMN), GLSZM_ZE: Zone Entropy (ZE), FO_MaxI: Maximum Intensity (MaxI), GLSZM_LAHGLE: Large Area High Gray Level Emphasis (LAHGLE), NGTDM_S: Strength (S), FO_MI: Mean Intensity (MI), GLDM_DV: Dependence Variance (DV), GLRLM_REn: Run Entropy (REn), GLCM_IDN: Inverse Difference Normalized (IDN), FO_RMS: Root Mean Square (RMS), GLSZM_ZP: Zone Percentage (ZP), NGTDM_Con: Contrast (Con), GLCM_Co: Contrast (Co), GLSZM_GLNN: Gray Level Non-Uniformity Normalized (GLNN), GLCM_Corr: Correlation (Corr), GLSZM_SAHGLE: Small Area High Gray Level Emphasis (SAHGLE), GLRLM_GLV: Gray Level Variance (GLV), GLCM_IV: Inverse Variance(IV), FO_V: Variance (V), FO_rMAD: Robust Mean Absolute Deviation (rMAD), GLRLM_GLNN: Gray Level Non-Uniformity Normalized (GLNN), NGTDM_B: Busyness (B), GLSZM_GLV: Gray Level Variance (GLV), GLCM_CS: Cluster Shade (CS), FO_IQR: Interquartile Range (IQR). Abbreviations: SHAP: SHapley Additive exPlanations, SL: Supervised learning, SSL: Semi-supervised learning, IDH: Isocitrate Dehydrogenase, GBM: Glioblastoma, RFs: Radiogenomics Features.

In the SSL setting, Fig. 4(ii) illustrates the average SHAP values for the same set of RFs, derived from the top 25 model-FSA/AEAs combinations. The SSL framework leverages unlabeled data through pseudo-labeling to enhance feature separability and model performance. Similar to the SL analysis, each row represents a selected RF, with SHAP values averaged for class 0 and class 1. Class 1 (Mutant IDH): Root Mean Square (FO_RMS, original, T1CE) emerged as the most discriminative feature, with a strong positive SHAP value of +0.887 for class 1 and a negative SHAP value of -0.275 for class 0. This indicates its significant role in favoring mutant IDH predictions while reducing confidence in wild-type predictions. Also, Class 0 (Wild-type IDH): Features such as Contrast (GLCM_co, wavelet LLL, T1CE) and Gray Level Non-Uniformity Normalized (GLSZM_GLNN, wavelet LLL, T1) showed negative SHAP contributions, suggesting they reduce model confidence for class 0 predictions. In conclusion, the SSL framework enhances the consistency and separability of key features, particularly texture- and zone-based RFs, by leveraging unlabeled data to refine decision boundaries. This results in improved model performance and interpretability compared to the SL setting. The SHAP-based analysis reveals distinct patterns in feature importance between SL and SSL settings. In SL, class 1 features, particularly wavelet-based textural features, dominate model predictions, highlighting their potential as biomarkers for IDH mutation. In SSL, the incorporation of unlabeled data amplifies the discriminative power of features like FO_RMS, suggesting that pseudo-labeling enhances the robustness of FSAs and model interpretability. These findings underscore the value of SSL in multi-center MRI-based GBM studies, where data heterogeneity and limited labeled samples are common challenges.

## 4. Discussion

GBM is the most aggressive primary brain tumor, and IDH mutation status remains a key prognostic biomarker: IDH-wildtype tumors follow a poorer clinical course than IDH-mutant gliomas [1, 2]. While MRI and radiogenomics provide a non-invasive pathway for IDH prediction, debate continues over the most informative imaging sequence [7, 8]. Advances in ML, particularly SSL, are shifting this paradigm by leveraging both labeled and unlabeled data to improve prediction accuracy in settings where annotated data are scarce. This multicenter study is novel in integrating SL and SSL frameworks with RFs extracted from multiple MRI sequences, systematically comparing individual and combined modalities, and embedding interpretability analysis through SHAP. To our knowledge, this is one of the first large-scale multicenter studies to evaluate the prediction of IDH mutation status with such a comprehensive design, addressing limitations of prior single-center, single-sequence, or purely supervised approaches.

Our results demonstrated that SSL consistently outperformed SL, yielding higher accuracy and greater robustness to dataset size variation. The best SL configuration (RFE + SVM) on the combined T1, T2, T1CE, and FLAIR sequences achieved a strong cross-validation accuracy of 0.92, whereas the SSL framework reached 0.93 and maintained stable performance even with limited labeled samples. SHAP-based analysis revealed that SSL amplified the discriminative power of features such as FO_RMS and wavelet-based textural descriptors, producing more consistent separation between IDH-mutant and wild-type tumors and enhancing biomarker reliability. Importantly, multi-sequence MRI (T1, T2, T1CE, FLAIR) achieved the highest diagnostic accuracy across both frameworks, while SSL markedly rescued weak single-sequence performance. Clinically, these results indicate that SSL reduces dependence on extensive annotation, strengthens underperforming modalities, and reinforces multi-sequence imaging as the most reliable approach for glioma subtyping and treatment planning.

Our findings align with emerging literature on SSL in biomedical prediction. In lung cancer, SSL significantly improved survival outcome prediction across multiple trajectories [36]. Another study showed that SSL incorporating unlabeled, diverse datasets—such as head and neck cancer alongside lung cancer—enhanced survival prediction in multi-fold cross-validation [36]. Similarly, SSL improved pathogenic variant prediction in Parkinson's disease, outperforming SL as a framework for genetic stratification [37]. These results support our conclusion that SSL can generalize across diseases, data types, and modalities. Salmanpour et al. (2025) reported SSL gains of up to 17% for CT-based lung cancer prognosis, even when only 10% of cases were labeled, underscoring its robustness and cost-effectiveness across multicenter cohorts [20]. Our study extends these insights to neuro-oncology, demonstrating that SSL maintains performance advantages in multicenter MRI-based molecular prediction.

The relationship between dataset size and model performance is critical across domains ranging from medical imaging to cancer diagnostics and even remote sensing, highlighting the universal importance of data efficiency. In our study, SSL achieved high accuracy (0.91–0.93) even with limited labeled data, while SL performance plateaued and showed diminishing returns as labeled samples increased. Similar patterns were observed by Al-Azzam et al. [38], who found that SSL achieved competitive accuracy (90–98%) with only half the labeled data. Ramezan et al. [39] further demonstrated that ensemble-based SL methods, such as Random Forest, remained robust under severe label reductions, while algorithms like SVM and neural networks were highly sensitive. Our findings corroborate these observations, as SSL stabilized fragile models (e.g., SVM), enabling them to achieve consistent performance with fewer annotations. Mechanistically, this suggests that SSL leverages unlabeled data to refine decision boundaries, reduce noise sensitivity, and mitigate overfitting risks, while ensemble-based SL methods may provide resilience in parallel under constrained conditions.

The integration of SSL with SHAP-based interpretability underscores the dual importance of accuracy and transparency. In our GBM analysis, SSL improved the discriminative capacity of features such as FO_RMS (T1CE) and wavelet-derived descriptors, surpassing the interpretability and stability observed in SL settings. This parallels the findings of Salmanpour et al. [20], where SSL amplified the prognostic value of texture- and zone-based features in CT lung cancer cohorts. Mechanistically, these results suggest that SSL enhances feature separability by reducing sensitivity to noise and leveraging pseudo-labeling to stabilize decision boundaries. This not only improves model accuracy but also supports reproducible biomarker discovery by highlighting biologically plausible, cross-center consistent features.

For translation into practice, SSL offers three major benefits. First, it reduces annotation demands, lowering the resource and time burden associated with manual labeling. Second, it improves the utility of weaker MRI sequences such as T2 and FLAIR, which are widely available in routine neuro-oncology but typically underperform when used alone. Third, SSL reinforces multimodal integration, confirming that no single sequence can consistently match the predictive power of combined data. T1CE alone performed nearly as well as multimodal integration, reflecting its biological relevance in capturing tumor vascularity and enhancement, but the added stability of combining multiple sequences confirms the value of fusion strategies. Clinically, these findings suggest that SSL-driven radiogenomics models can support earlier IDH prediction when biopsy is contraindicated, when molecular testing is delayed, or in resource-limited settings, thereby guiding treatment planning and patient counseling.

This study has several limitations. It's retrospective design and inter-site protocol variability may have introduced confounding factors, although SSL's consistent performance across centers suggests resilience to such heterogeneity. While external validation was included, prospective evaluation in independent cohorts is essential for clinical deployment. Additionally, this work focused exclusively on IDH mutation; extending SSL to other biomarkers such as MGMT methylation and 1p/19q codeletion will be critical for comprehensive molecular profiling. Future research should also explore SSL across multi-omics datasets—integrating imaging, genomic, and pathology features—and assess real-time integration of SSL models into decision support systems.

## 5. Conclusion

Our multicenter study shows that SSL reliably outperforms SL for non-invasive prediction of IDH mutation status in glioblastoma. By leveraging unlabeled data, SSL improved accuracy, generalization, and feature stability, offering greater robustness in data-limited and heterogeneous settings. SHAP-based analysis further demonstrated that SSL enhanced the discriminative power of key RFs, reinforcing interpretability and supporting biomarker discovery. Clinically, SSL rescued the diagnostic value of weaker sequences such as T2 and confirmed multi-sequence integration (T1, T2, T1CE, FLAIR) as the most reliable strategy for glioma molecular stratification. These results establish SSL as a scalable and label-efficient framework that reduces annotation demands while increasing the reliability of imaging biomarkers. In summary, SSL advances radiogenomics toward clinically actionable decision support in neuro-oncology, providing a pathway to more accessible, interpretable, and precise precision-care tools.

## DECLARATIONS

**Supplemental Files, Data, and Code Availability.** All code (including prediction and dimension reduction algorithms) is publicly shared at: https://github.com/MohammadRSalmanpour/Semi-Supervised-Radiomics-for-Glioblastoma-IDH-Mutation. Supplemental File 1 contains the machine learning performance results for Supervised Learning (SL) using FLAIR. Sheet 1 ("Selected_Features") represents the top 10 features selected by each feature selector or attribute extraction method. Sheet 2 ("Best_Parameters") represents the optimal hyperparameters of the applied machine learning algorithms. Sheet 3 ("Aggregated_Results") represents the aggregated validation and test performance metrics, including Accuracy, Precision, AUC, F1-score, Recall, and Specificity. Sheet 4 ("Aggregated_Std_Results") represents the corresponding standard deviation values. Supplemental File 2 contains the machine learning performance results for SL using combined sequences. (Sheets 1–4 as described above). Supplemental File 3 contains the machine learning performance results for SL using T1. (Sheets 1–4 as described above). Supplemental File 4 contains the machine learning performance results for SL using T1CE. (Sheets 1–4 as described above). Supplemental File 5 contains the machine learning performance results for SL using T2. (Sheets 1–4 as described above). Supplemental File 6 contains the machine learning performance results for Semi-Supervised Learning (SSL) using FLAIR. (Sheets 1–4 as described above). Supplemental File 7 contains the machine learning performance results for SSL using combined sequences. (Sheets 1–4 as described above). Supplemental File 8 contains the machine learning performance results for SSL using T1. (Sheets 1–4 as described above). Supplemental File 9 contains the machine learning performance results for SSL using T1CE. (Sheets 1–4 as described above). Supplemental File 10 contains the machine learning performance results for SSL using T2. (Sheets 1–4 as described above).

**Acknowledgements.** This study was supported by the Virtual Collaboration Group (VirCollab, www.vircollab.com) and the Technological Virtual Collaboration (TECVICO CORP.) based in Vancouver, Canada. We gratefully acknowledge funding from the Natural Sciences and Engineering Research Council of Canada (NSERC) Discovery Horizons Grant DH-2025-00119.

**Conflict of Interest.** Drs. Mohammad R. Salmanpour and Mehrdad Oveisi are affiliated with TECVICO Corp. The remaining authors declare no relevant conflicts of interest.